# Generation of Chinese-characters-like modes with transverse mode locked lasers

Zilong Zhang[1,2,*], Yuan Gao[1,2], Changming Zhao[1,2]

1 School of Optics and Photonics, Beijing Institute of Technology, 5 South Zhongguancun Street, 100081 Beijing, China

2 Key Laboratory of Photoelectronic Imaging Technology and System, Ministry of Education of People's Republic of China

**Abstract:** Spontaneous transverse mode locking to generate optical vortex arrays is already widely studied for the frequency degenerated transverse modes. While, the mode locking between transverse modes in different orders (non-frequency degenerated families) to achieve spatial stationary beam patterns is rarely reported. The theory of transverse mode locking is discussed. Concepts of general transverse mode locking (G-TML) and restricted transverse mode locking (R-TML) are mentioned. It's experimentally shown that R-TML is possible in microchip cavities with high nonlinearity. More interestingly, some Chinese characters like laser modes can be generated by microchip lasers with this special transverse mode locking effect. And various experimental results and corresponding simulations are presented.

## 1. Introduction

Since the coming out of laser sixty years ago, the development of all characters of laser is such rapid to make it a very important light source for so many areas from scientific researches to industrial applications. Almost all the characters can be concluded to temporal, spatial or spectral domain. The laser properties in temporal and spectral domains are paid lots of attentions in the past decades, while spatial properties of laser beam seem to be relatively neglected. With the spring up of researches on vortex beams and vector beams, laser spatial properties of phases and polarizations by external modulations gets more attentions. Although intra-cavity generations of vortex or vector beams are presented by some works [1-6], the focus on the internally generated laser pattern properties based on cavity transverse modes still remains less.

Hermit-Gaussian or Laguerre-Gaussian modes are eigenmodes of a laser cavity based on classical laser theory. Ince-Gaussain modes were also proved to be eigenmdoes for off-axis pumped laser cavities [7]. Besides the eigenmode states output from laser cavity, transverse mode locked (TML) states can also be generated with cavity intrinsic nonlinearity [8-10]. Beam patterns generated by TML effect usually carries optical vortices arrays [8-13], which are of great values in lots applications. However, these studies only focused on the phase locking of transverse modes in frequency-degenerated families. Recently, both phase and frequency locking of transverse modes in different orders were proved possible by us [14]. The combination of transverse modes in different orders brings a new vision to us both in understanding the laser physics and recognizing the vast possibilities for laser beam patterns.

In this paper, transverse mode locking effect is studied in theory and experiment. And the concept of restricted transverse mode locking (R-TML) is mentioned and discussed. We also present an interesting phenomenon of generating Chinese characters like beam patterns by the restricted transverse mode locking effect. Laser beam patterns like some Chinese characters which have the structure similar to crystal lattice are obtained. These beam patterns are all spatio-temporally stable ones, maintaining the relative intensity distributions unchanged with time and propagation. Theoretical simulations show good agreement with the experimentally measured beam patterns, which not only proves the correctness of the theoretical adjustments, but also helps us to understand the exact formation information of the Chinese characters like laser modes. This technology may be of great value in laser display, particle manipulations and free space communications.

## 2. Theories of transverse mode locking

The phenomenon of transverse mode locking (TML) is firstly reported in 1968 [15,16]. Several studies on TML were also reported after that [17-19]. Similar to longitudinal mode locking, TML was usually considered as a locking of phase difference. Optical frequencies difference of the transverse modes was not cared. However, if the frequencies of transverse modes were not the same, the obtained beam patterns were not stationary with propagation [20-22]. It's not easy for one to use these beams in real applications.

Usually, it's considered that the transverse modes in different orders possess different optical frequencies in a laser cavity. And the total electric field of superposition of transverse modes in $0$ to $n$th order takes the form,

$$E_{tot}(x,y,z) = \sum_{0}^{n} A_n f_n(x,y,0) \exp\left(i\omega_n t + i\frac{\omega_n}{c}\frac{x^2+y^2}{R_n(z)} - i(n+1)\psi(z) + i\varphi_{n,0}\right) \quad (1)$$

Here, $A_n$ is the weight of each mode, $f_n(x,y,0)$ is the normalized transverse electric field amplitude distribution function at $z=0$ plane, and $\exp(\cdots)$ is the total phase of the $n$th transverse mode. $\omega_n$ is the optical frequency, $R(z)$ is the radius of curvature of the wave front, $\psi(z)$ is the Gouy phase, and $\varphi_{n,0}$ is a constant phase of the $n$th transverse mode.

For a set of frequency degenerated modes, Eq.(1) could be rewritten as below,

$$E_{tot}(x,y,z) = \exp\left(i\omega_n t + i\frac{\omega_n}{c}\frac{x^2+y^2}{R_n(z)} - i(n+1)\psi(z)\right) \cdot \sum_{0}^{m} A_m f_m(x,y,0)\exp(i\varphi_{m,0}). \quad (2)$$

Here, we spouse $m$ transverse modes are all in the $n$th order. As these modes are frequency degenerated, the spatial and temporal evolutions of their phases are in the same value.

While, for the transverse modes which are not in the same frequency degenerated families, cooperative frequency locking effect is possible to pull the frequencies of these modes to an averaged one with the help of nonlinearity of the laser cavity [22], and then the total electric field can be expressed as,

$$E_{tot}(x,y,z) = \exp\left(i\overline{\omega} t + i\frac{\overline{\omega}}{c}\frac{x^2+y^2}{\overline{R(z)}} - i\overline{q\psi}(z)\right) \cdot \sum_{0}^{n} A_n f_n(x,y,0)\exp(i\varphi_{n,0}). \quad (3)$$

Here, $\overline{\omega}$ is the averaged optical frequency, whose derivation is as below,

$$\begin{cases} \overline{\omega} = \omega_0 + \dfrac{\sum_{0}^{n} A_n \Delta\omega_n}{\sum_{0}^{n} A_n}, \\ \Delta\omega_n = \omega_n - \omega_0 = \dfrac{c(k_x^2 + k_y^2)}{2k_z} \end{cases} \quad (4)$$

Here, $\omega_0$ is the optical frequency of the fundamental transverse mode, $\Delta\omega_n$ is the frequency spacing between the $n$th mode and the fundamental mode, $c$ is velocity of light, and $k_x=\pi n/l_x$, $k_y=\pi n/l_x$, $k_z=2\pi/\lambda$. $l_x$ and $l_y$ are the sizes of the cavity in $x$ and $y$ direction, respectively. Along with the locking of frequencies, the parameters of $R(z)$ and index of $\psi(z)$ that instead wave propagations should also be an averaged one to help with the locking of the total phases.

Once the variable phase item in Eq.(1) is split out, the electric field's transverse intensity distribution $E_{tot}(x,y)$ can be determined by the integration item as in Eq.(2) or Eq.(3). The three items in the integration function determine the mode locked beam patterns, which usually contains spatio-temporally stationary optical vortices arrays. And different combination of these three items will lead to quite different results. The formation of $f_n(x,y)$ is one of the expressions of basic laser modes (Hermit-Gaussian or Laguerre Gaussian). The integration of these basic laser modes will show locking properties, and the constant phases $\varphi_{n,0}$ count much for the exact intensity distribution of the obtained beam patterns.

For these obtained TML modes, three special characters should be mentioned. Firstly, the obtained mode is a single frequency one. The slightly frequency difference (usually in hundreds of MHz or several GHz level [23]) between the locked modes is eliminated by the frequency coupling effect. Secondly, the TML modes are linearly polarized, which also comes with the coupling effect. Actually, the second character is also related to the first one, due to different polarizations will lead to multi frequency oscillations, considering the anisotropy of an active gain medium. And thirdly, the energy distributions of the output beam patterns defer to minimizing the energy of the laser cavity [24,25].

To the condition of Eq.(2), it's the locking of frequency degenerated transverse modes. And it is usually easy to be observed in Class B lasers. We call this locking phenomenon the general transverse mode locking (G-TML). To the condition of Eq.(3), the locked transverse modes are in different frequency families and are locked to an averaged value by cooperative interactions, we call it the restricted transverse mode locking (R-TML). In a laser cavity with high nonlinearity, we can achieve the R-TML effect [14].

## 3. Experiments and Results

The laser cavities we used in the experiment are microchips composed of dual-medium, with one as gain medium and the other as high nonlinear medium. The microchip cavities can achieve a high Fresnel number pump condition easily, which is essential for the R-TML. And the microchip cavity can offer single frequency operation for each transverse mode, which is also very important for the R-TML state. As multi longitudinal mode operation may increase the difficulty of phase locking between transverse modes. What's more, due to the unlocked phases between the longitudinal modes, the obtained TML beam patterns will not be pure.

The experimental diagram of our work is shown in figure 1. Variations of output beam patterns are achieved by adjusting of the pumping parameters, such as pump powers, focused beam waist in chip or the incident angles. We have already obtained some rarely reported beam patterns based on the R-TML effect in Ref.[14]. While, the Chinese characters like beam patterns are obtained this time. This property of the obtained beam patterns due to a common ground between the TML modes and the Chinese characters. It's that the beams generated by TML has a large tendency to form the structure with optical vortices arrays, which is quite similar to a usually used basic structure emanated by cross-hatching in Chinese characters. There are also some differences, the contour profile of a Chinese character is usually a quadrate one, while that of the TML mode prefers to be a circular one due to the circular pump beam profile. However, under some cases of superposition of different transverse modes, the contour profile can be reshaped by phase singularities and be a quadrate one, thus resulting in a Chinese character like one.

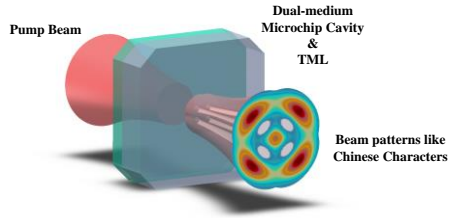

FIG.1 Diagram of the generation of Chinese characters like beam patterns from a dual-medium microchip cavity based on TML effect. The laser beam is a single frequency one with linear polarization.

Here, we would like to show the experimental results firstly and then give its generation principle by numerical simulations. One of the most similar characters we achieved in the experiment is the "田" (read as "Tian" in Chinese, and it means the Farmland) beam pattern. This pattern matches the optical vortices array property quite well, as its four optical vortices arranged with a four-quadrant construction like in figure 2. Figure 2(a)-2(c) show three beam patterns which are all quite similar to the character "田" (Tian) just with slight differences. The insets in each of the three subfigures are three different typefaces of "田". These beam patterns were obtained under similar pump conditions, which were 6.5 W 808 nm laser power be focused into the Nd:YAG+LTO$_3$ dual-medium microchip cavity by near normal incidence. The slight differences between the three patterns come from a tiny tuning of the pump beam's incident angle. And by tuning the incident angle relatively largely, in several degrees. Some more beam patterns can be obtained shown as in figure 3. The beam patterns in figure 3(a)-3(f) are similar to the Chinese characters "工" (Gong, means the Work), "王" (Wang, means the King), "日" (Ri, means the Sun), "回" (Hui, means Coming back), "三" (San, means Three), and "吕" (Lv, is a Family name), respectively. Here, for figure 3(c) and 3(d), the figures should be rotated by 90 degrees to match the corresponding characters.

The beam patterns in figure 2 and 3 are all spatio-temporally stable ones emanated by restricted TML effect, which are rarely seen before. None of them are basic modes in laser cavity. Though the pattern in figure 3(e) looks quite similar to the mode TEM$_{02}$, it's also a hybrid one.

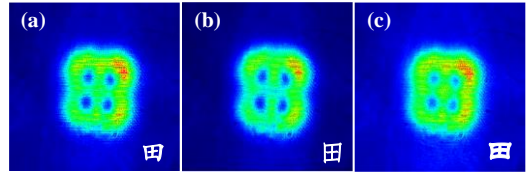

FIG. 2. The experimentally obtained beam patterns which are similar to the Chinese character "田 (Tian)".

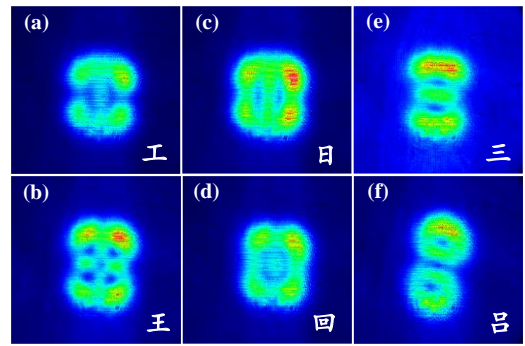

FIG. 3. Some more beam patterns which are also quite similar to Chinese characters.

As discussed in the second section, complex beam patterns can be generated by the TML effect. However, investigations are still needed to prove that these patterns are results of TML by what kinds of basic modes and locked phases. While, it was a really hard work to match out these beam patterns with so many basic modes with variable intensities and different phase locking possibilities. Here, we show some numerical simulation cases corresponding with the experimentally obtained beam patterns to prove the correctness of the theory. Figure 4(a)-(d) show the numerically simulated beam patterns that are coincident with figure 2(a), figure 3(a), 3(b) and 3(e), respectively. The subfigures in the first to third lines are beam intensities, phases and the locking methods of basic modes (the minus sign means a phase difference of π). Compared with the beam patterns in figure 2 and 3, we could see that the similarity between them is really high.

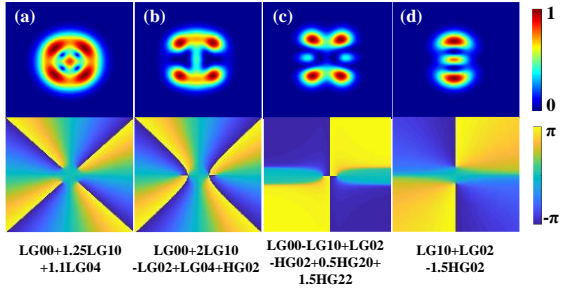

FIG. 4. Numerical simulations of the TML states of some basic modes to show good coincidence with the experimentally measured ones.

With another kind of microchip cavity (Nd:YAG+$Cr^{4+}$:YAG, with 300 μm in thickness of Nd:YAG, and 200 μm in thickness of $Cr^{4+}$:YAG), similar laser modes were also achieved by TML effect. Some experimentally measured beam patterns are shown in figure 5, where (a) to (c) are quite similar to some Chinese characters, while (d) to (f) are similar to just a portion of some Chinese characters. For figure 5 (a) and (b), they are very similar to the ones in figure 3 (f) and figure 2, respectively. It fingers out that, the generation of Chinese characters similar laser modes is not a special case with a certain cavity, but a relatively common case with the microchip cavities that are easy to obtain the restricted TML states. For figure 5 (c), the beam patterns needs an anti-clockwise rotation with an angle of 45 degree to match the character of '小' (Xiao, which means small). And for figure 5 (f), it is a simple and clear evidence of the restricted TML states. The simulated beam phases and patterns of figure 5 (f) are shown in figure 5 (g) and (h), respectively. With simulation, it can be seen that the laser mode in figure 5 (f) is a locked state of $HG_{02}$ mode and $HG_{04}$ mode with a phase difference of half π.

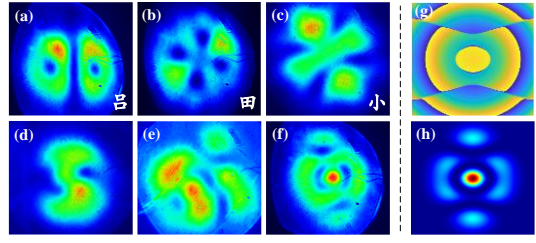

FIG. 5. Beam patterns obtained with the composed Nd:YAG+$Cr^{4+}$:YAG microchip cavity.

And there is another possibility of the laser beams. As most of the Chinese characters are composed of some simple characters or parts, we could also achieve some complex characters with the directly obtained ones by combining them together. Here, we would like to show a composed character of '福' (Fu) as an example, shown in figure 5 (composed of figure 2(a)). The character is very similar to a kind of Chinese penmanship. '福' means good luck in Chinese. As this paper is prepared during the serious pneumonia epidemic period all over the world. We would like to wish the world a good luck!

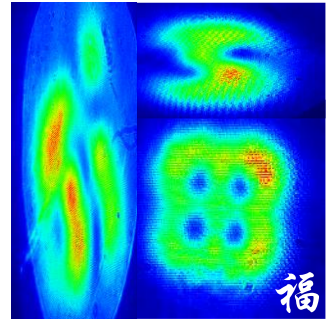

FIG. 6. The '福' character composed of figure 2(a), 5(d) and 5(e). Here, we reshaped figure 5(d) and 5(e). Though this picture is just an editing of the figures, it is also possible to achieve it by experimental method, if we got three these micro cavity systems.

## 4. Conclusions

In conclusion, we theoretically discussed the transverse mode locking effect and gave the concept of restricted transverse mode locking (R-TML) effect. It's experimentally show that Chinese characters like beam patterns can be generated by transverse mode locking effect, especially with the locking of different transverse mode orders, namely the restricted TML effect. Numerical simulations show good agreement with the experimental results.

**Acknowledgements**

This research is funded by the National Natural Science Foundation of China, grant number 61805013.